\documentclass[twocolumn]{revtex4}
\usepackage{graphicx,amsmath,amssymb,mathrsfs}
\usepackage{epsfig}

\begin{document}

\title{Statistics of Lagrangian quantum turbulence}

\author{Christian Beck and Shihan Miah}

\affiliation{Queen Mary University of London, School of Mathematical Sciences, Mile End Road, London E1 4NS, UK}

\begin{abstract}
We consider
the dynamics of small tracer particles in turbulent quantum fluids.
The complicated interaction processes of vortex filaments, the
quantum constraints on vorticity and the varying influence of both
the superfluid and the normal fluid on the tracer particle
effectively lead to a superstatistical
Langevin-like model that in a certain approximation can be solved
analytically.
An analytic expression for the PDF of velocity $v$ of the tracer particle is derived
that exhibits not only
the experimentally observed $v^{-3}$ tails but also the correct behavior
near the center of the distribution, in excellent agreement
with experimental measurements and numerical simulations. Our
results
are universal and do not
depend on details of the quantum fluid.
\end{abstract}
\maketitle

Quantum turbulence is a phenomenon of utmost interest in current
fluid mechanics research \cite{ref11,ref16,ref17,ref40,ref41,ref42,ref44,ref42a,ref77,ref15}. The turbulent behavior of a quantum
liquid such as $\,^4$He is very different from classical turbulence
since vortices are quantized. This means the circulation cannot take
arbitrary values as in classical turbulence, and there is also no
viscous diffusion of vorticity as in classical turbulence.
Recent measurements \cite{ref11} and simulations \cite{ref16,ref17} have shown
that this has profound influence on various measurable observables, most notably
the velocity distributions of small test particles embedded in
the turbulent  flow. Whereas for classical turbulence there is near-Gaussian
behavior \cite{ref46,ref47,ref48}, one typically observes power laws
for quantum turbulence \cite{ref11,ref16,ref17,ref66}.

The velocity statistics 
%of test particles in quantum turbulence flow 
has been
subject of several recent papers.
Paoletti \textit{et al} \cite{ref11} conducted a
 seminal experiment using solid hydrogen tracers in turbulent superfluid $^4$He and found that the distribution of velocity components $v_i$ of the
 tracer particles exhibits a power law $p(v_i)\propto v_i^{-3}$ distribution
 for large values of $v_i$.
Similar results were confirmed by White \textit{et al} \cite{ref17}. They performed numerical simulations of quantum turbulence in a trapped Bose-Einstein condensate by calculating solutions of the  Gross-Pitaevski equation.
The associated PDF of each velocity component $v_i (i=x,y,z)$ was computed directly and  it was confirmed that the velocity statistics is non-Gaussian and obeys a power-law distribution $p(v_i) \propto v_i^{-b}$ with $-3.6<b<-3.3$.
In the following, for ease of notation,
we often suppress the index $i$.
%\begin{equation*}
%i\hbar\frac{\partial \psi}{\partial t}=-\frac{\hbar^2}{2 m'} \nabla^2 \psi+V_0 |\psi|^2 \psi-E_0 \psi
%\end{equation*}
Adachi \textit{et al} \cite{ref16} numerically computed the velocity field of
 a superflow by calculating the Biot-Savart velocity induced by vortex filaments in steady counterflow turbulence. They found that the resulting PDF exhibits a near-Gaussian distribution in the low-velocity region whereas a power-law $p(v)\propto v^{-3}$ is observed in the high velocity region.

Apparently there is clear evidence from numerical and experimental approaches that power laws in the velocity
statistics are highly relevant in quantum turbulence, and that
typically the observed power law exponent is close to 3. What is missing so far, however, is
a theory of {\em Lagrangian quantum turbulence}, by which we mean a theory that
consistently describes the dynamics of tracer particles 
of a given size embedded in the quantum turbulent flow,
which would be 
the proper theoretical tool to explain the observed velocity distributions.
Whereas Lagrangian turbulence is a well-established subject area for classical
turbulence \cite{ref3,ref2,ref4,ref56,ref57,ref58,ref59}, very little is known for the quantum case.

In this paper we will introduce a simple but powerful dynamical model
of the dynamics of a tracer particle embedded in a quantum
liquid. This model will be based on a superstatistical
stochastic differential equation. The superstatistics concept,
introduced in \cite{ref12}, has proved to be a very powerful method for
modeling a variety of complex systems 
\cite{bcs,hasegawa,thurner,garcia}, including
driven nonequilibrium situations
and classical hydrodynamic turbulence \cite{ref1,ref2,ref3,ref4}.

Here, for the first time, we apply this concept to quantum turbulence.
The result is a dynamical theory that quite precisely reproduces the
observed velocity statistics
in quantum turbulence and that also allows for
some analytic predictions. In particular, the power law exponent $-3$
follows from our theoretical consideration in a natural way, and moreover a universal prediction
for the entire shape of the velocity distribution is obtained, which is in excellent agreement with
experimental measurements.
%We also obtain
%the correct energy spectrum.

Let us denote the velocity of a Lagrangian tracer particle embedded in the quantum liquid  by $\boldsymbol{v}(t)$. We start from a simple local dynamics which will later
be extended to a superstatistical model.
Consider a linear stochastic differential equation
of the form
\begin{equation}
\dot {\boldsymbol{v}}(t)=-\Gamma\boldsymbol{v}(t)+\Sigma\boldsymbol{L}(t)
\label{eq1}
\end{equation}
Here $\boldsymbol{L}(t)$ is a rapidly fluctuating stochastic process representing rapid forces in the quantum liquid on a fast time scale, and $\Gamma$ and $\Sigma$ are $3\times3$ matrices.
The above equation simply says that locally a tracer particle is driven by chaotic
forces $\boldsymbol{L}(t)$ from the turbulent flow and at the same time
there are damping processes, described by $\Gamma$. Since the chaotic forces act rapidly
we approximate $\boldsymbol{L}(t)$ by Gaussian white noise. $\Gamma$ and $ \Sigma$ are matrix-valued stochastic processes which evolve on a much larger time scale than $\boldsymbol{L}(t)$. The particle is driven by a mixture of normal and superfluid,
and depending on which component dominates, the effective friction 
described by $\Gamma$ will be very different.

A characteristic property of quantum turbulence is a spatio-temporally
varying vorticity field represented by 1-dimensional topological
vortices that reconnect and merge at random moments of time.
A test particle may rotate for a short
while around a local unit vector $\boldsymbol{e}$ whose direction
will be a random
variable, describing a given vortex filament in the
 quantum liquid. Hence, as a special case of eq.~(\ref{eq1}) we may
consider the local dynamics
\begin{equation}
 \dot {\boldsymbol{v}}=-\gamma(t)\boldsymbol{v}+\omega \big[\boldsymbol e(t) \times \boldsymbol v\big] +\sigma\boldsymbol{L}(t)
\label{eq2}
\end{equation}
We assume that the damping constant $\gamma$ and the noise strength $\sigma$ are functions of $t$, and so
 is $\omega$ and the direction of $\boldsymbol{e}$. The second term on the right hand side of eq.(\ref{eq2}) represent the rotational movement of the particle around the vortex filament. The unit vector $\boldsymbol{ e}$ and the noise strength $\sigma$ evolve stochastically on a large time scale $T_{\boldsymbol e}$ and $T_\sigma$ respectively.\\

A special coordinate system would be $\boldsymbol{e}=(0,0,1)$, then
$\boldsymbol{ e} \times \boldsymbol v=(-v_y, v_x, 0)$
and the velocity components of the particle satisfy
\begin{equation}
\begin{split}
 \dot v_x=-\gamma {v}_x-\omega {v}_y+\sigma {L}_x(t)\\
\dot v_y=-\gamma v_y+\omega v_x+\sigma L_y(t)\\
\dot v_z=-\gamma v_z+\sigma L_z(t)\\
\end{split}
\label{eq3}
\end{equation}
If we introduce a complex variable $z$ by defining  $z=v_x+iv_y$, then the $(x,y)$-dynamics can be written as
\begin{equation}
 \dot z=\dot v_x+i\dot v_y=(-\gamma+i\omega)z+\sigma (L_x+i L_y)
\label{eq4}
\end{equation}
 Forming the average $\langle \cdots \rangle$ over all
 realizations of the
 noise $\boldsymbol{L} (t)$ one obtains on a time scale
 where $\gamma$ and $\omega$ are sufficiently constant
\begin{equation}
\langle z(t)\rangle =z(0)e^{-\gamma t}(\cos(\omega t)+i\sin(\omega t))
\label{eq6}
\end{equation}
which is just damped spiraling motion around a local unit vector
with frequency $\omega$. We remind the reader
that the basic idea of the superstatistics approach is to regard the
{\em parameters} of a local stochastic differential equation as
random variables as well \cite{ref1}. This means both $\gamma$ and
$\omega$ can take on very different values during time evolution,
and so can the direction of $\boldsymbol{e}$.
A very small $\gamma$ corresponds to nearly undamped motion
for a limited amount of time.
A very small $\omega$ corresponds to almost no rotation, i.e. straight
movement for a limited
 amount of time. All these cases are included as possible local dynamics
 and averaged over in the superstatistical approach.

In a quantum turbulent flow, the superfluid component flows without dissipation while being subject to certain quantum mechanical constraints. These quantum restrictions imply that the typical form of rotational motion allowed in the superfluid component is
 in the form of a thin vortex line, whose circulation around its core is quantized rather than arbitrary as in classical fluids. The  magnitude of the  velocity field of the fluid particle at
distance $r$ from the core of the vortex filament is given by \cite{ref16}\\
\begin{equation}
 v=|\boldsymbol v|=\frac{\kappa}{2\pi r}
\label{eqq}
\end{equation}
            where $\kappa =\frac{h}{m}\approx 9.97\times 10^{-4} cm^2/s$ is the quantum of circulation, $h$ is Planck's constant and $m$ is the mass of the fluid atom, in
            our case helium.\\

 If the
 tracer particle comes close to
  a vortex filament, it will
  typically follow a circular path around the vortex filament, with 
$ v=\frac{2\pi r}{ T}=\frac{\kappa}{2\pi r}$,
 where $T$ is the period of one rotation. Note that the angular frequency
 entering eq.~(\ref{eq4}) is
 thus $\omega=\frac{2\pi}{T}=\frac{\kappa}{2\pi r^2}$.

                             For an ordinary spherical Brownian particle
 in a viscous liquid one has
 constant damping due to Stoke's law:\\
\begin{equation}
 \gamma=\frac{6\pi\nu\rho a}{M} \label{e8}
\end{equation}
  Here $\nu$ is the kinematic viscosity of the liquid, $\rho$ is the fluid density, $M$ is the mass  and $a$ the radius of the tracer particle.\\

For quantum turbulence, the effective dissipation acting on the tracer particle
is influenced by many competing effects, and it fluctuates strongly depending
on whether the particle is close to a vortex filament or not. Far away from
a vortex filament, the movement will be dominated by Brownian motion similar as in
a normal liquid, whereas close to a vortex filament the movement will
be very rapid and almost friction free, dominated by the superfluid.

To take into account the fact that the effective friction in eq.~(\ref{eq2})
is fluctuating, we may write quite generally
\begin{equation}
\gamma=\frac{1}{L^2} \nu \sum_{i=1}^n  X_i^2 \label{e9}
\end{equation}
where $L$ is a characteristic length scale
 and the $X_i$ are dimensionless random variables that evolve in time and space.
 We have squared the random variables because for
 physical reasons $\gamma$ must always be positive,
 though values close to 0 are possible.
 $n$ denotes the number of degree of freedoms that influence the fluctuating effective
 friction. Of course, the simplest model is to assume that
 the $X_i$ are a rescaled sum of many microscopic random variables that
 act almost independently. Thus the Central Limit Theorem suggests to assume
 that the $X_i$ are Gaussian random variables.
 %We will be able to test
 %this hypothesis with experimental data later.
 %, and will also come
 %up with a prediction for the value for $n$.

%In classical fluids, turbulence occurs when the Reynolds number $Re$ is relatively high, that means the kinematic viscosity is small because the Reynolds number $Re=\frac{UD}{\nu}$ is the ratio of the magnitudes of the inertial and viscous forces acting on the fluid. As the pure superfluid is non-viscous, the Reynolds number in superfluid flow is infinite nominally and the dissipation exists only because the organised kinetic energy is converted into disorganised sound energy \cite{ref15}. \\

The quantum mechanical constraint given in Eq. (\ref{eqq}) tells us that the
average rotational velocity of the tracer particle is very high near the vortex core (for small
distance $r$). Therefore, the effective viscosity $\gamma$
acting on the tracer particle in eq.~(\ref{eq2}) is small
if the particle is very near to the vortex core.
This means that
$\sum_i^n X_i^2$ is small.
On the other hand, if the test particle is very far from the vortex filament, then $\gamma$ is
large and the friction effects are strong, mainly due to the normal fluid component.
This suggests the physical interpretation that the $X_i$'s may just be
       identified with the perpendicular distances of the test  particle from the nearest vortex filament. The vortex
       filaments themselves of course
       evolve in a highly complicated stochastic way. Since only the distance perpendicular to the nearest vortex filament is relevant,
       for a 3-dimensional quantum liquid we have $n=2$, that is 
two degrees of freedom. The distance $r$ of the test particle from the
 vortex filament becomes a random variable given by
\begin{equation}
 r^2=\left(X_1^2+ X_2^2\right) L^2, \label{e10}
\end{equation}
 where again $L$ is  a suitable spatial scale introduced for dimensional reasons.

 We may estimate this length scale $L$ as follows: For large distances $r$, of the order of
 average vortex filament distance $d$ in the turbulent flow, the tracer particle follows
 nearly normal type of Brownian motion, with Stokes law (\ref{e8}) valid in good approximation.
 Putting $r=\frac{d}{2}$ into eq.~(\ref{e10}), 
 (\ref{e9}) and (\ref{e8}) one arrives at
 the following estimate for the length scale $L$:
 \begin{equation}
 L = \left( \frac{M d^2}{24 \pi \rho a} \right)^{\frac{1}{4}}
\end{equation}
%Note that this result is independent of the kinematic viscosity $\nu$. Our typical
%length scale $L$ grows proportional to
%the square root of the average vortex filament distance $d$.
Clearly, our model requires small particles with $a << d$,
if larger scales $a>d$ are probed, one just gets ordinary Brownian
motion with Gaussian behavior \cite{salort,baggaley}.

The velocity distribution of the 
small
tracer particle in the quantum turbulent flow described by eq.(\ref{eq2}) can now be 
calculated by using standard techniques
of superstatistics
\cite{ref4}.
We first assume, for simplicity, a constant $\gamma$ and define the parameter $\beta :=\frac{2\gamma}{\sigma^2}$, which in equilibrium statistical mechanics corresponds to the inverse temperature, whereas here
it is more a measure of distance from the nearest vortex filament.
%\footnote{In our mathematical treatment
%this is just a formal analogy, $\beta$ describes the strength of local
%chaotic forces in the turbulent flow, not the physical temperature of the quantum liquid.}.
On time scales $t$ satisfying $\gamma^{-1}\ll t\ll T_\sigma$ the stationary distribution of the tracer particle described by Eq.(\ref{eq2}) for fixed $\beta=\frac{2\gamma}{\sigma^2}$ is given by the Gaussian distribution\\
\begin{equation}
p(v|\beta)=\sqrt{\frac{\beta}{2\pi}}e^{-\frac{1}{2}\beta v^2},
 \label{eq10}
\end{equation}
assuming uniform distribution of the random vectors $\boldsymbol{e}$.
The situation becomes different for fluctuating $\beta$, that is, if one allows the parameters $\gamma$ (or $\sigma$) in Eq.(\ref{eq2}) to be varying as well. Assuming
that $\hat{X}_1, \ldots \hat{X}_n$ are independent Gaussian random variables, the resulting distribution of $\beta=\sum_{i=1}^n \hat{X}_i^2$ is  a $\chi^2$ distribution of degree $n$, i.e.

\begin{equation}
 f(\beta)=\frac{1}{\Gamma (\frac{n}{2})}\left(\frac{n}{2\beta_0}\right)^{\frac{n}{2}} \beta^{\frac{n}{2}-1} e^{-\frac{n\beta}{2\beta_0}}
\label{eq11}
\end{equation}
The average of the fluctuating $\beta$ is given by
\begin{equation}
 \langle \beta\rangle=n\langle \hat{X_i}^2\rangle=\int_0^\infty \beta f(\beta)=\beta_0
\end{equation}
and the variance by
\begin{equation}
 \langle \beta^2\rangle-\beta_0^2=\frac{2}{n}\beta_0^2
\end{equation}

The probability density to observe the velocity
 $v$ of the test particle for any value of  $\beta$  is given by the marginal probability $p(v)$ as follows
\begin{equation}
 p(v)=\!\! \! \int_0^\infty f(\beta) p(v|\beta) d\beta\\
\label{eq14}
\end{equation}
 Substituting $p(v|\beta)$ and $f(\beta)$ from Eq.(\ref{eq10}) and Eq.(\ref{eq11}) into Eq.(\ref{eq14}), we obtain after a short calculation

\begin{equation}
 p(v)
=\frac{\Gamma(\frac{n}{2}+\frac{1}{2})}{\Gamma(\frac{n}{2})} \left(\frac{\beta_0}{\pi n}\right)^\frac{1}{2} \frac{1}{\left(1+\frac{\beta_0}{n} v^2\right)^{\frac{n}{2}+\frac{1}{2}}}
\label{eq16}
\end{equation}
These types of distributions play an important role in
$q$-generalized versions of statistical mechanics \cite{tsallis},
with the entropic index $q$ related to the parameter $n$ by $q=1+\frac{2}{n+1}$.

As we mentioned earlier the velocity of the tracer particle depends on the perpendicular distance between the particle and the
nearest evolving (and sometimes merging) vortex filament. Therefore, the relevant degrees of freedom are $n=2$ 
for 3-dimensional quantum turbulence. By substituting $n=2$  in Eq.(\ref{eq16}) one obtains

\begin{equation}
 p({v})=
\frac{\sqrt{\beta_0}}{(2+\beta_0 v^2)^\frac{3}{2}}
\label{eq177}
\end{equation}
Clearly, for large $v$ this implies power-law tails
\begin{equation}
p({v})\propto {v}^{-3}.                                                                                                              \end{equation}
The remarkable result, however, is that we do not only get the power law tails
but a concrete prediction for the entire shape of the probability distribution,
including the region near the maximum.

The probability distribution of kinetic
energy $E$ can be calculated from Eq.(\ref{eq16}) by using a
simple transformation of random variables.
For a particle of unit mass $E=g(v)=\frac{1}{2} v^2$, hence $v=g^{-1}(E)= \sqrt{2E}$ and
\begin{equation}
p_E(E)dE=p_v(v)\; 2dv,
\end{equation}
the factor 2 coming from the fact that there are two solutions $\pm v$ for the
same energy $E$. This leads to the probability distribution of energy
\begin{equation}
\begin{split}
 p_E(E)=2 p_v\left(g^{-1}(E)\right)\left|\frac{dg^{-1}(E)}{dE}\right|\\
= \sqrt{2} \frac{\Gamma(\frac{n}{2}+\frac{1}{2})}{\Gamma(\frac{n}{2})} \left(\frac{\beta_0}{\pi n}\right)^\frac{1}{2} \frac{1}{(1+\frac{2\beta_0}{n} E)^{\frac{n}{2}+\frac{1}{2}}} \frac{1}{\sqrt{E}}%(1+\frac{2\beta_0}{n} v) %^
\end{split}
\label{eqq188}
\end{equation}
For $n=2$ this predicts power law tails proportional to $E^{-2}$ for large $E$.

So far our model was based on a situation where the average velocity $v$ of the
particle is zero. Of course, in experiments
there is often a drift velocity in the system
that gives a non-zero mean velocity $c$ to the test particle.
In this case one has to replace $v$ by $v-c$ in the
model equations we derived so far, and for $n=2$ one ends up with
\begin{equation}
 p(v)=\frac{\sqrt{\beta_0}}{\left(2+\beta_0 (v-c)^2\right)^\frac{3}{2}}
\label{eq188}
\end{equation}

      Let us now compare
      our model prediction
      with the experimental data obtained by Paoletti et al. \cite{ref11}.
Fig.~1 shows the experimental data for both velocity
components $v_x$ and $v_z$, and a fit by our analytic formula.
An excellent fit is obtained.
\begin{figure}
\includegraphics[width=3.5in]{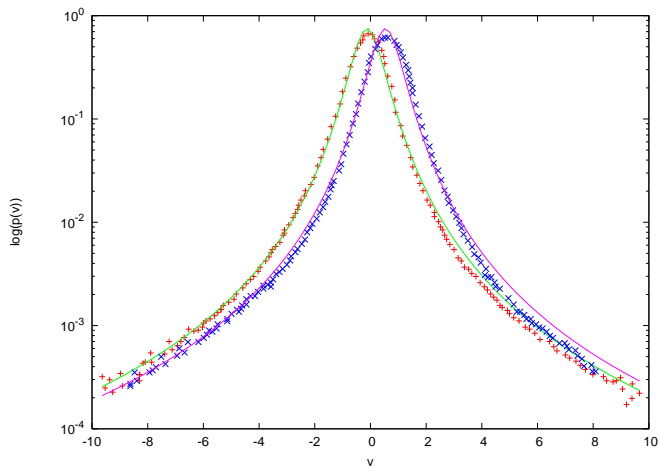}
\caption{Experimental data of Paoletti et al \cite{ref11} and a fit using eq.~(\ref{eq188})
  with variance parameter $\beta_0=4.5$ and  $c=-0.12$ for $v_x$,
  respectively $c=0.54$ for $v_z$}
\label{fig1}
\end{figure}
It is remarkable that the fit is not only correctly producing the power law tails
but also the vicinity
of the maximum. To illustrate this, Fig.~2 shows the same data in a linear
plot.

\begin{figure}
\includegraphics[width=3.5in]{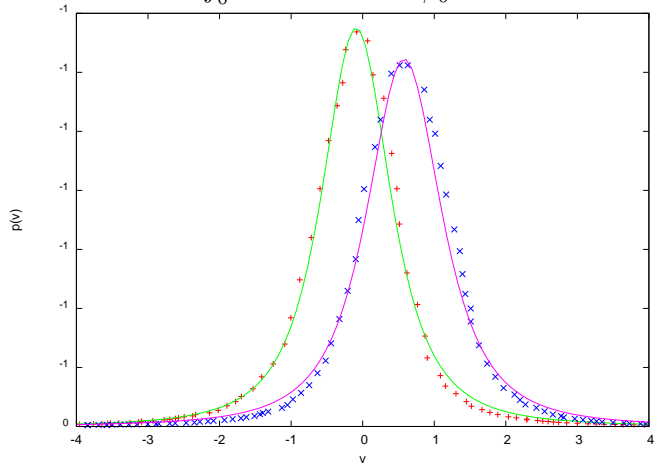}
\caption{Same as Fig.~1 but a linear scale is chosen, which emphasizes the
vicinity of the maximum. The variance parameter is slightly adjusted to
obtain optimum coincidence in the central region.}
\label{fig2}
\end{figure}

Let us mention that our model directly predicts the power law exponent $-3$
in a universal way. The value $-3$ is a consequence of the fact that vortex filaments are
thin 1-dimensional structures embedded into 3-dimensional space, thus leading to $n=2$
in eq.~(\ref{e10}).
 Our model also correctly reproduces
the $E^{-2}$ tails of the energy spectrum observed by Paoletti et al \cite{ref11}.

Finally, we can also predicts the value $\beta_0 \simeq 4.5$ of the variance parameter
to be used in eq.~(\ref{eq188}).
So see this, let us recall that Paoletti et al., in their experiment \cite{ref11},
rescaled their measured velocity data to variance 1.
In these units their maximum velocity measured
was $v_{max} \simeq 10$ (see Fig.~1). Strictly speaking, the
variance does not exist for any distribution that decays as $v^{-3}$ for large $|v|$,
but what exists is of course the variance as calculated for a given experimental cutoff $v_{max}$.
From
\begin{equation}
1= \langle v^2 \rangle \simeq 2 \int_0^{v_{max}}p(v)v^2dv
\simeq \frac{2}{\beta_0} \log v_{max}
\end{equation}
we obtain the predicted value $\beta_0 \simeq 2 \log v_{max} \approx 4.6$, in agreement with what yields the optimum
fit in Fig.~1. Thus, besides the (nonuniversal) systematic drift velocity $c$, all relevant parameters 
are predicted from first principles.

To conclude,
in this paper we have developed
a superstatistical dynamical model of Lagrangian quantum turbulence. This model predicts
that
   the velocity statistics of 
small tracer particles in a quantum turbulent flow obey a
   power law distribution $p(v)\propto v^{-3}$ and the distribution of energy follows a power law as well,
    i.e. $p(E)\propto E^{-2}$. These results are in excellent agreement with Paolleti
    et al.'s measurements \cite{ref11} as well as with the numerical results obtained by other authors \cite{ref16, ref17}.
Our theory provides a universal prediction given by (\ref{eq188}) for
 both the center and the tail parts of the velocity distribution. The underlying stochastic model
      arises quite naturally out of the fact that small
tracer particles see fluctuating effective frictions, depending on the distance to the nearest vortex filament.
%The result is very different from classical turbulence, where velocities are in good approximation Gaussian and where velocity {\em %differences} obey lognormal superstatistics \cite{ref4}.

%\clearpage


\begin{thebibliography}{h!}
\bibliographystyle{ama}

\bibitem{ref11}\textsc{M. Paoletti, M. Fisher, D. Lathrop, K. Sreenivasan}, Phys. Rev. Lett. \textbf{101} 154501 (2008).
\bibitem{ref17}\textsc{A.C. White, C.F. Barenghi, N.P. Proukakis, A.J. Youd, and D.H. Wacks}, Phy. Rev. Lett. \textbf{104} 075301 (2010).
\bibitem{ref16}\textsc{H. Adachi, M. Tsubota}, Phys. Rev. B \textbf{83} 132503 (2011).
\bibitem{ref40} \textsc{T. Araki, M. Tsubota, and S.K. Nemirovskii}, Phys. Rev. Lett. \textbf{89} 145301 (2002).
\bibitem{ref41} \textsc{C. Nore, M. Abid, and M.E. Brachet}, Phys. Rev. Lett. \textbf{78} 3896 (1997).
\bibitem{ref42} \textsc{M. Kobayashi and M. Tsubota}, Phys. Rev. Lett. \textbf{94} 065302 (2005).
\bibitem{ref44} \textsc{N.G. Parker and C.S. Adams}, Phys. Rev. Lett. \textbf{95} 145301 (2005).
\bibitem{ref42a} \textsc{M. Kobayashi and M. Tsubota}, Phys. Rev. Lett. \textbf{76} 045603 (2007).
\bibitem{ref77} \textsc{D.R. Poole, C.F. Barenghi, Y.A. Sergeev, and W.F. Vinen}, Phys. Rev. B. \textbf{71}, 064514 (2005).
\bibitem{ref15}\textsc{C.F. Barenghi}, Physica D \textbf{237}, 2195 (2008).
\bibitem{ref46} \textsc{A. Vincent and M. Meneguzzi}, J. Fluid Mech. \textbf{225} 1 (1991).
\bibitem{ref47} \textsc{A. Noullez, G. Wallace, W. Lempert, R.B. Miles and U. Frisch}, J. Fluid Mech. \textbf{339} 287 (1997).
\bibitem{ref48} \textsc{T. Gotoh, D. Fukayama and T. Nakano}, Phys. Fluids \textbf{14} 1065 (2002).
\bibitem{ref66} \textsc{I.A. Min, I. Mezic, and A. Leonard}, Phys. Fluids \textbf{8}, 1169 (1996).
\bibitem{ref56} \textsc{A. La Porta, G. A. Voth, A. M. Crawford, J. Alexander, and E. Bodenschatz}, Nature \textbf{409} 1017 (2001).
\bibitem{ref57} \textsc{A. M. Reynolds, N. Mordant, A. M. Crawford, and E. Bodenschatz}, New Journ. Phys. \textbf{7} 58 (2005).
\bibitem{ref58} \textsc{N. Mordant, P. Metz, O. Michel, and J.-F. Pinton}, Phys. Rev. Lett. \textbf{87} 214501 (2001).
\bibitem{ref59} \textsc{N. Mordant, E. Leveque, and J.-F. Pinton}, New Journ. Phys. \textbf{6} 116 (2004).
\bibitem{ref2}\textsc{C. Beck}, Europhys. Lett. \textbf{64} 151 (2003).
\bibitem{ref3}\textsc{A.M. Reynolds}, Phys. Rev. Lett. \textbf{91} 084503 (2003).
\bibitem{ref4}\textsc{C. Beck}, Phys. Rev. Lett. \textbf{98} 064502 (2007).
\bibitem{ref12}\textsc{C. Beck, E.G.D. Cohen}, Physica A \textbf{322} 267 (2003).
\bibitem{bcs}\textsc{C. Beck, H.L. Swinney, and E.G.D. Cohen}, Phys. Rev. E \textbf{72} 056133 (2005).
\bibitem{hasegawa}\textsc{H. Hasegawa}, Phys. Rev. E \textbf{83} 021104 (2011).
\bibitem{thurner} \textsc{R. Hanel, S. Thurner, and M. Gell-Mann}, PNAS \textbf{108} 6390 (2011).
\bibitem{garcia} \textsc{V. Garcia-Morales, K. Krischer}, PNAS \textbf{108} 19535 (2011).
%\bibitem{ref32} \textsc{C.Beck}, Physica A \textbf{331}, 173 (2004).
%\bibitem{ref29} \textsc{C.Beck}, Physica D \textbf{193}, 195 (2004).
%\bibitem{ref9}\textsc{L.Tisza}, Nature. \textbf{141}, 913(1938).
%\bibitem{ref10}\textsc{L. Landau}, Phys. Rev.Lett. \textbf{60}, 356-58(1941).
\bibitem{ref1}\textsc{C.Beck}, Phys. Rev. Lett. \textbf{87} 180601 (2001).
\bibitem{salort} \textsc{J. Salort, B. Chabaud, E. L\'ev$\hat{e}$que, and P.-E.
Roche}, Europhys. Lett. \textbf{97} 34006 (2012). 
\bibitem{baggaley} \textsc{A.W. Baggaley and C.F. Barenghi},
arXiv:1110.5767v2.
\bibitem{tsallis} \textsc{C. Tsallis}, J. Stat. Phys. \textbf{52} 479 (1988).
\end{thebibliography}
\end{document}